\documentclass{elsart}
\usepackage{natbib}

\usepackage[dvips]{graphics,color,graphicx}
\usepackage{multicol}

\begin{document}
\runauthor{Zach,Murphy,Marriott,Boyd}
\begin{frontmatter}
\title{OPTIMIZATION OF THE DESIGN OF OMNIS, THE OBSERVATORY OF MUTLIFLAVOR
NEUTRINOS FROM SUPERNOVAE}

\author[The Ohio State University]{J.~J.~Zach}
\author[The Ohio State University]{A~St\,J.~Murphy}
\author[The Ohio State University]{D.~Marriott}
\author[The Ohio State University,Astronomy]{R.~N.~Boyd}
\address[The Ohio State University]{Department of Physics, Smith Laboratory, 
174 W. 18th Ave., Columbus, OH, 43210}
\address[Astronomy]{Department of Astronomy, 
140 W. 18th Avenue, Columbus, OH, 43210}
\begin{abstract}
A Monte Carlo code has been developed to simulate the operation of the planned 
detectors in OMNIS, a supernova neutrino observatory. OMNIS will detect neutrinos 
originating from a core collapse supernova by the detection of spalled neutrons from 
Pb- or Fe-nuclei. This might be accomplished using Gd-loaded liquid scintillator. 
Results for the optimum configuration for such modules with respect to both neutron detection
efficiency and cost efficiency are presented. Careful consideration has been given to the 
expected levels of radioactive backgrounds and their effects. The results show that 
the amount of data to be processed by a software trigger can be reduced to 
the $<10~kHz$ region and a neutron, once produced in the detector, can be detected 
and identified with an efficiency of $>30\%$.
\end{abstract}
\end{frontmatter}

\section{Introduction}

The electron-antineutrinos observed by the Kamiokande \cite{kamiokande} and 
IMB \cite{IMB} laboratories from supernova 1987-A confirmed that neutrinos 
play an important role in core collapse supernovae. The vast majority of the 
gravitational binding energy of the protoneutron star is radiated away by neutrinos. 
The observed $\bar{\nu _e}$ were detected over a few seconds, much longer than 
the weak interaction time-scales which are thought to be required to produce them. 
This confirmed the main features of the standard model of core-collapse 
supernovae~\cite{9606035} and suggests that the behavior of the neutrino opacity in 
matter beyond nuclear densities is understood \cite{9410229}. As has been shown in 
numerous supernova simulations (see, e.g. \cite{mezzacappaliebendoerfer2000}), the shock 
that will produce the supernova stalls, thus disfavoring a prompt mechanism 
\cite{baroncooperstein90}. Although there are many uncertainties in the 
calculations, indications that the shock has to be revived by neutrinos are strong \cite{bethe931}. 
\par
An increasing body of data from the atmospheric neutrino experiments at Super-Kamiokande 
\cite{boezio99}, IMB \cite{beckerszendy}, Soudan2 \cite{Allison99} and Kamiokande \cite{kamiokandeatm}, 
exhibit the disappearance of $\mu$-neutrinos \cite{0002147}. Solar neutrino experiments such 
as Homestake \cite{homestake}, Kamiokande \cite{kamiokande2}, GALLEX \cite{Wanninger99}, 
SAGE \cite{Abdurashitov99} and Super-Kamiokande \cite{superksolar} also suggest a flux of 
$\nu _e$'s inconsistent with the Bahcall-Pinsonneault standard solar model. The most favored
solution to these discrepancies is an extension of the standard model of particle physics to one
in which the neutrinos have mass and are allowed to oscillate between flavors via the vacuum and/or
the Mikheyev-Smirnov-Wolfenstein (MSW-) oscillation mechanisms \cite{9906469}. Direct 
measurements of the neutrino masses \cite{9902462} yield upper limits of $m_{\nu _e} \le 
3~eV$ (90$\%$CL), $m_{\nu _{\mu }} \le 170~keV$ (90$\%$CL) and $m_{\nu _{\tau }} \le 18.2~MeV$ 
(95$\%$CL). However, the tremendous neutrino luminosities of supernovae provide a unique 
scenario for extremely long baseline measurements, allowing a vastly increased region of 
neutrino mixing parameter space to be probed, and placing more stringent limits than can
otherwise be achieved on neutrino masses and mixing angles.
\par
The astrophysical r-process may also depend critically on the properties of neutrinos. 
The leading candidate for its site is the low-density, high-entropy bubble 
of a core-collapse supernova, created in the trail of the outgoing shockwave by material 
lifted off the surface of the protoneutron star and heated by neutrinos \cite{woosley94, 
Takahashi94}. The details of the interaction between the neutrinos and matter in the bubble 
are highly nonlinear, and may require several types of MSW-type oscillations \cite{9910175, 
mclaughlinfetter99}. 

\section{Motivation and Properties of OMNIS}

\subsection{The Role of OMNIS in the Supernova Neutrino Observatory Community}

An observation of the fluxes of different neutrino flavors is vital for the 
understanding of core-collapse supernovae and helpful for the investigation of new 
physics beyond the standard model. In this context, the need for a large neutral-current
neutrino detector has become increasingly obvious in recent years \cite{cline94}. Accordingly, 
OMNIS is being designed to provide a much larger sample of $\mu$- and $\tau$-neutrino 
events than would be provided by other detectors, allowing a significantly more detailed 
measurement of the neutrino spectra for different flavors, resolved in time, and the
stellar conditions that produced them. Given the planned quantities of iron and lead, and 
with the detection strategy suggested here, a typical supernova at the center of the Galaxy 
would lead to about 2000 observed events in OMNIS \cite{nnn99} (assuming no oscillations), 
predominantly from {\em neutral-current} reactions. These events will provide a window for 
looking well beyond the peak of the supernova neutrino burst to study the evolution at times up to 
$\approx10~seconds$ after the start of the pulse, a potentially important diagnostic 
of the cooling mechanism. A few thousand events are also adequate for accurate neutrino mass 
measurements. 
\par
Most present neutrino detectors, such as Super-Kamiokande, LVD, MACRO and AMANDA,
are primarily sensitive to charged-current channels, involving mainly the reaction of
$\bar{\nu _e}$ with protons in water or mineral oil. Super-Kamiokande will also see 
some $e$-neutrinos and some $\mu$- and $\tau$-neutrinos through the detection
of $\gamma$-rays produced in the neutral-current channels $^{16}O(\nu ,\nu ^{\prime } 
p/n \gamma)$, as suggested by Langanke and Vogel \cite{superknc}. The total 
number of detected neutral-current events due to $\mu$- and $\tau$-neutrinos for a 
core-collapse supernova at a distance of $8~kpc$ has been estimated to be around 
300 or 560, depending upon the temperature and chemical potential in the neutrino Fermi-Dirac 
distribution assumed. Recent changes to the data acquisition that result in a lower detection 
threshold might improve this to around 1000 events~\cite{vagins}. However, this yield comes
from a single reaction threshold and so the unknown neutrino energy distribution cannot be 
inferred. SNO, which is designed around a tank of heavy water, would be expected to produce around
750 break-up events of deuterium following neutrino neutral-current scattering for a supernova 
at the same distance \cite{beacomvogel}. It is on the other hand scheduled for a lifetime at the 
low edge of the mean time between Galactic supernovae (10-30 years, \cite{0006015}), even though 
SNO's scheduled time has been extended from two to ten years. 
\par
 It has been shown \cite{beacomvogel} that, based on time-of-flight information, 
Super-Kamiokande and SNO will have resolutions of $m_{\nu _{\mu /\tau }} \approx50~eV$ and
$m_{\nu _{\mu /\tau }} \approx30~eV$, respectively, although again the improved statistics
due to a lower detection threshold will improve considerably the value for Super-Kamiokande. 
Due to both the greater number of neutral-current events detected, and to its superior timing 
capability, OMNIS is expected to have a still better mass resolution of around~$20~eV$.
\par
A particularly interesting scenario is that in which a Galactic core-collapse 
supernova proceeds to a black hole. This should produce an abrupt termination of 
neutrino flux as well as several remarkable measurements. Firstly, the signal preceding 
the cut-off, and when the signal occurs may depend on the protoneutron star evolution, and 
hence the equation of state~\cite{baumgarte}. Secondly, as the black hole expands outward
from the center of the star it will envelope successive neutrinospheres~\cite{baumgarte}. 
One might therefore expect the $\nu_{\mu /\tau }$-flux to be terminated earlier than the 
$\bar{\nu _e}$-flux which would in turn occur before the termination of the $\nu _e$-flux. 
The difference in time between these terminations has been estimated to be of order $1~ms$ 
\cite{baumgarte}. If the supernova is close enough, OMNIS would be able to observe these 
differences, and thus could infer the structure of the neutrinospheres. Lastly, the 
termination of the luminosity provides a well defined feature that would provide an opportunity 
for time-of-flight mass measurements of the neutrino mass with unparalleled accuracy, down to 
limits of $2~eV$ for $\bar{\nu _e}$ in Super-Kamiokande and as low as $4~eV$ for $\mu$- and 
$\tau$-neutrinos from OMNIS \cite{0006015}.

\subsection{Expected Neutrino Signal from a Core-Collapse Supernova}

In order to make estimates of the response of the OMNIS detectors to a supernova, we have 
included standard expectations for the incident neutrino intensities, energy distributions and
interaction cross sections. The bulk of the neutrino luminosity from a core-collapse supernova 
is emitted during the Kelvin-Helmholtz cooling phase, which is marked by an exponentially 
decaying neutrino flux with a time constant of $\sim$3~$seconds$ \cite {ponsreddy}. Each flavor 
is described by a Fermi-Dirac distribution with a chemical potential at or near zero. At the relevant
energies, the $\mu$- and $\tau$-neutrinos only interact via neutral-current processes, and
therefore decouple deeper within the core (at a smaller neutrinosphere radius, the term being 
defined as the radius from the center at which neutrinos decouple from the dense matter), 
leaving them with the highest energies, estimated to be around $<E_{\nu_{\mu /\tau }}> 
\approx25~MeV$ 
or $T \approx8~MeV$. Electron-neutrinos decouple at a larger neutrinosphere radius, with the 
net numerical superiority of neutrons to protons resulting in the $e$-antineutrinos 
having higher energy than the $e$-neutrinos ($<E_{\nu_{\bar{e}}}>\approx16~MeV$ or $T\approx5~MeV$ 
versus $<E_{\nu_{e}}>\approx11~MeV$ or $T\approx3.5~MeV$ \cite{qian93}). 

\subsection{Neutrino Cross Sections Relevant to OMNIS}

OMNIS will consist of several $\frac{1}{2}~kT$ modules of lead and iron, which will serve as
target materials. Some of the incoming neutrinos from a supernova will undergo charged-current 
or neutral-current interactions with the composite nuclei, resulting in neutron emission. The cross 
sections for such reactions depend on the neutrino energy, the Q-value for the reaction 
and the availability of suitable energy levels in the target nucleus. There are no available 
experimental data for the cross sections governing these reactions, with the exception of that 
for $^{56}Fe(\nu _e,e^-)^{56}Co$ measured by the KARMEN collaboration \cite{karmen}, but theoretical 
studies have been conducted by two groups \cite{FHM,nEspec}. Despite qualitative agreement between 
the two groups, some uncertainty remains, due to the largely unknown location of the Gamow-Teller 
resonance relative to the neutron separation energies. These separation energies are 
$10.67~MeV$ for $^{56}Fe$ and $7.19~MeV$ for $^{208}Pb$ for single-neutron emission and $19.78~MeV$ 
and $13.31~MeV$, respectively, for double-neutron emission. The single-neutron channels in lead 
and in iron and the two-neutron channel in lead constitute three observables closely related
to the neutral-current thresholds that can be used to deduce the relevant contributions of the 
various neutrino flavors via the different energy dependencies of their cross sections. Both SNO 
and Super-Kamiokande will provide additional reaction thresholds and their data could be used in 
collaboration with those from OMNIS to map out the neutrino distributions.

\subsection{Expected Number of Events in OMNIS}

All event numbers presented in this work are based on a ``standard'' supernova in the center of 
the Galaxy (distance = $8~kpc$, $E_{tot}=3\times 10^{53}~ergs$). The number of neutrons liberated per $kT$
in OMNIS modules of lead or iron is indicated in table \ref{prodratetable}. The cross sections for 
lead were taken from \cite{FHM} and for iron from \cite{nEspec}. 

\begin{table}
 \caption{Comparison for single- and double-neutron events liberated from iron and lead for 
the different reaction channels (per $kT$) 
\label{prodratetable}}
\begin{center}
\begin{tabular}{|l|r|r|r|r|r|r|}
\hline
 Material, Event Type & CC-$\nu _e$ & CC-$\bar{\nu }_e$ & NC-$\nu _e$ & NC-$\bar{\nu }_e$ & NC-$\nu _x$ & total\\
 \hline\hline
 Pb, single-n, no osc. & 59 & 0 & 8 & 37 & 677 & 781 \\
 \hline
 Pb, double-n, no osc. & 26 & 0 & 0 & 1 & 20 & 47 \\
 \hline
 Fe, single-n, no osc. & 4 & 5 & 2 & 6 & 146 & 163 \\
 \hline
\end{tabular}
\end{center}
\end{table}

\subsection{Influence of Neutrino Oscillations on the OMNIS Yields}

\begin{figure}[t]
{\centering
{\includegraphics[width=7cm, angle=90, clip=true, trim=0 0 0 0]{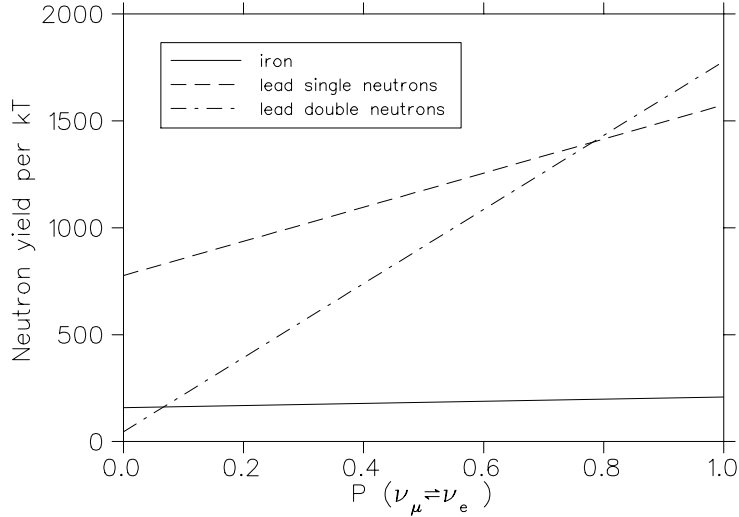}}
\caption{Neutron Yield per $kT$ Versus Oscillation Probability $\nu _{\mu } \leftrightarrow \nu _{e}$. \label{mueosc}}
}
\end{figure}

\begin{figure}[t]
{\centering
{\includegraphics[width=7cm, angle=90, clip=true, trim=0 0 0 0]{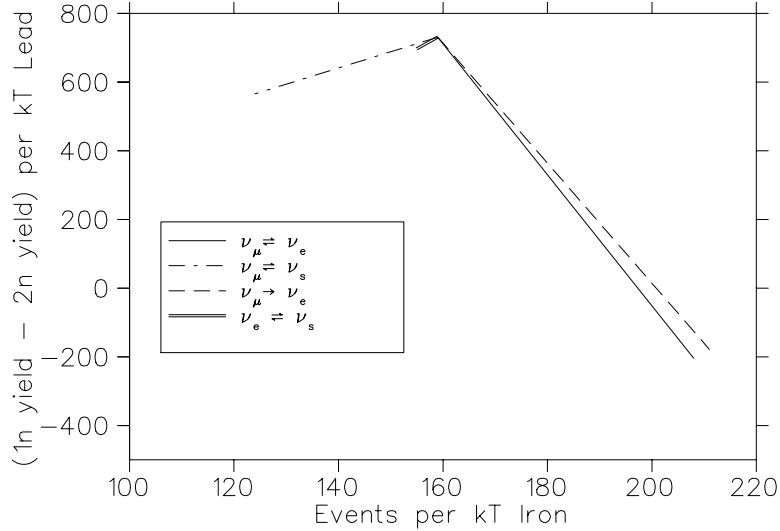}}
\caption{Single-Neutron Yields in the Pb and Fe modules (each per $kT$) for different Oscillation Schemes. \label{sensitivity}}
}
\end{figure}

One possible scenario is the oscillation between $\mu$-~or $\tau$-neutrinos and 
electron-neutrinos. The resulting higher energy for the latter, combined with the large 
charged-current cross section, leads to a significant increase in the overall yield in both 
the iron and the lead and to a dramatically increased number of double-neutron events in 
lead (see also table \ref{osctable}, which shows the result for an oscillation of all
$\mu$-neutrinos into $e$-neutrinos and all $e$-neutrinos into $\mu$-neutrinos). The latter channel 
therefore serves as a sensitive neutrino thermometer and/or indicator for certain types of 
neutrino oscillations. Figure~\ref{mueosc} shows the expected number of liberated neutrons 
per $kT$ as a function of $P$, the chance that a neutrino emitted from the supernova 
as a $\mu$-neutrino has oscillated into an electron-neutrino and vice versa. The ability of 
OMNIS to discriminate between possible oscillation scenarios is illustrated in fig. \ref{sensitivity}, 
where the difference between the number of single-neutron emission events and the number of 
two-neutron emission events observed in lead, is plotted against the number of neutron emission 
events in iron. This demonstrates the additional sensitivity that OMNIS has to oscillations 
involving sterile neutrinos.

\begin{table}
 \caption{Comparison for single- and double-neutron events liberated from iron and lead
for the different reaction channels (per $kT$) with full oscillation $\nu _{\mu } \leftrightarrow 
\nu _e$.\label{osctable}}
\begin{center}
\begin{tabular}{|l|r|r|r|r|r|r|}
\hline
Material, Event Type & CC-$\nu _e$ & CC-$\bar{\nu }_e$ & NC-$\nu _e$ & NC-$\bar{\nu }_e$ & NC-$\nu _x$ & total\\
 \hline\hline
 Pb, single-n & 826 & 0 & 184 & 35 & 516 & 1563 \\
 \hline
 Pb, double-n & 1852 & 0 & 6 & 1 & 15 & 1874 \\
 \hline 
 Fe, single-n & 57 & 5 & 37 & 6 & 112 & 217 \\
 \hline
\end{tabular}
\end{center}
\end{table}


\section{Features of the OMNIS detector}

\subsection{The OMNISita Test Facility}

OMNIS will consist of approximately $4~kT$ of iron and $8~kT$ of lead modules. It will be built 
in the Waste Isolation Pilot Plant (WIPP) near Carlsbad, NM, which has been shown, given its 
background radiation and available infrastructure, to be a highly desirable site~\cite{wipp}. 
At the time of this paper, a test module, OMNISita~\cite{omnisita}, is being constructed at the 
proposed site in order to test the predictions of the Monte Carlo code presented here, to have a test 
facility for the components of the OMNIS detector, and for further measurements of the 
radioactive and cosmic ray backgrounds.

\subsection{Gd loaded Liquid Scintillator for Neutron Detection}

Once neutrons are emitted, they interact with the neutron source material (Fe 
or Pb) and the neutron detector, assumed in the present study to be organic 
liquid scintillator. This serves three functions: neutron moderator, sink and detector. 
The use of Gd in liquid organic scintillators is well established 
\cite{psmith}, it has the highest known neutron capture cross section~\cite{Gdxsecs} 
and its solubility is high enough to make it the preferred neutron sink at this 
time. The possible alternatives Li and B have not been found to be soluble in 
organic liquids in sufficiently high concentrations and there are no known organic 
molecules containing Li or B that have a light yield similar to Gd loaded 
scintillator. 

\subsection{Particle Propagation and Event Identification}

\begin{figure*}
{\centering
\includegraphics[width=10cm]{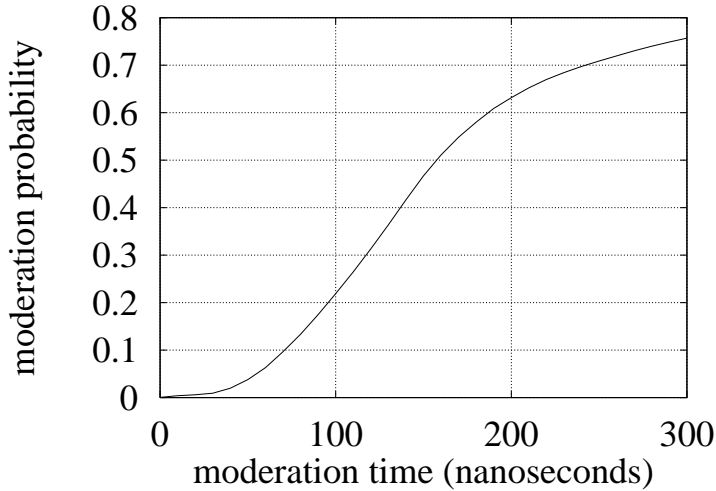}
\caption{Neutron Moderation Time-scale\label{moderation}}
}
\end{figure*}

\begin{figure*}
{\centering
\includegraphics[width=10cm]{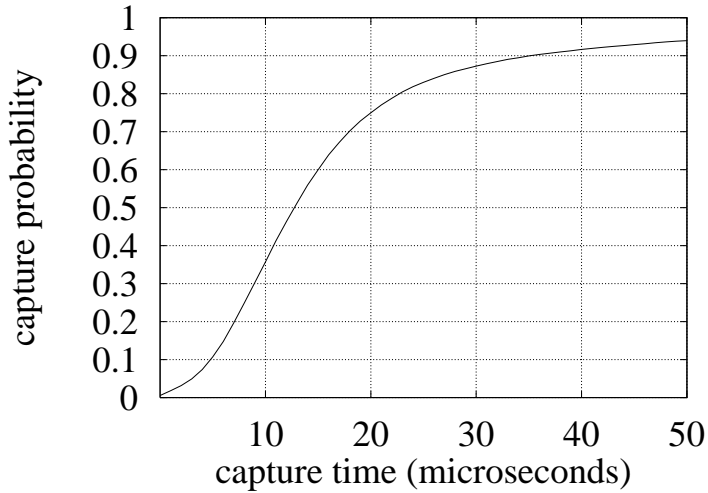}
\caption{Neutron Capture Time-scale\label{capture}}
}
\end{figure*}

The neutrons produced by the neutrino interactions are expected to have initial 
energies of $\sim1~MeV$ \cite{nEspec}, hence moderation will primarily occur through elastic 
scattering off the protons in the scintillator. Our calculations show that the 
neutrons lose nearly all of their kinetic energy within $\propto100~ns$ (see fig.
\ref{moderation}) after emission, resulting in the production of a significant 
number of scintillation photons. With a $0.1\%$ loading of Gd in the scintillator, capture 
occurs after $\approx30~\mu s$, see also figure \ref{capture}, predominantly on $^{155/157}Gd$. 
Figures \ref{moderation} and \ref{capture} were generated with the optimized design for the 
lead modules presented below, but these time-scales are not very dependent on the detector 
material and geometric configuration used. Further, the capture time scales are consistent with 
a recent study~\cite{nnn99, alexconf} performed using a prototype scintillator vessel and with 
similar Monte Carlo studies conducted in the past \cite{trzcinski96, trzcinski99}. To enable 
discrimination against cosmic rays and $\gamma$-rays produced by the decay of radioactive 
impurities in the lead or iron or other detector components, a double-pulse technique with a time 
window between prompt and capture signals of $\approx50~\mu s$ will be used. 
The characteristic second pulse, caused by the emission of $\gamma$-rays following capture on 
Gd or H in the scintillator vessels, is used to identify the neutron pulse. The time of neutron 
liberation is then known within a few $100~ns$. Although pulse shape discrimination might 
be considered as a means to identify the neutron induced events, this would be impractical 
for detectors as large as those required for a facility of the size of OMNIS. 

\subsection{Design Constraints}
\label{constraints}

\begin{figure*}
{\centering
\includegraphics[height=15cm, width=13cm, angle=270, clip=true, trim=50 50 0 0]{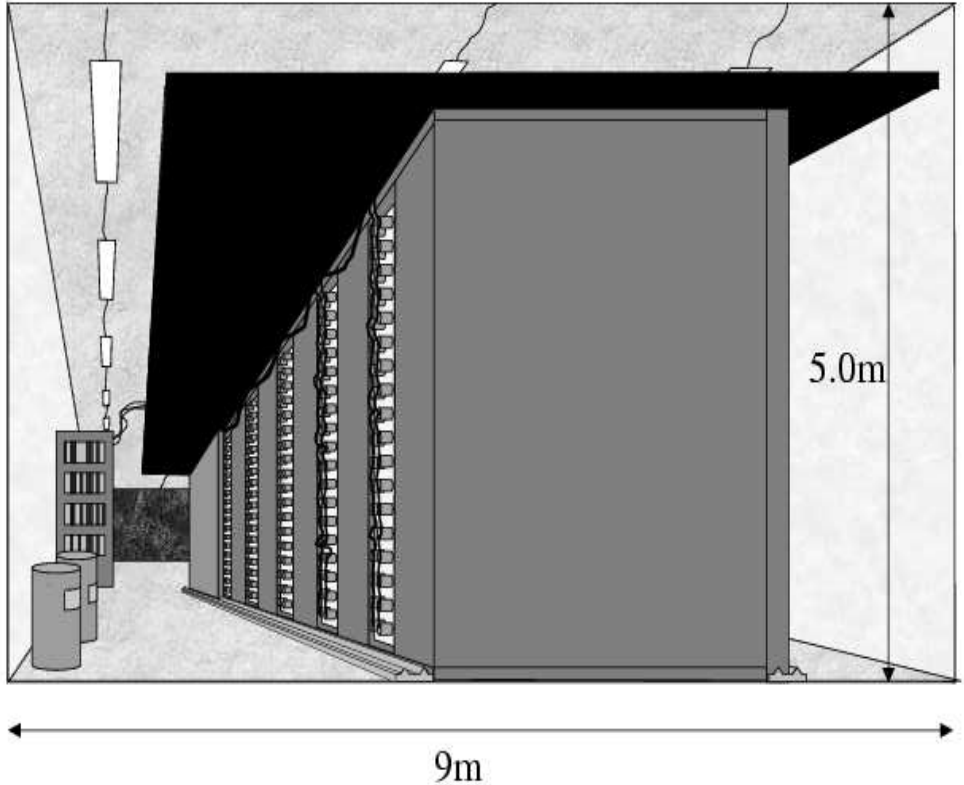}
\caption{Front View of an OMNIS Lead Module\label{f1}}
}
\end{figure*}

\begin{figure*}
{\centering
\includegraphics[height=15cm, angle=270]{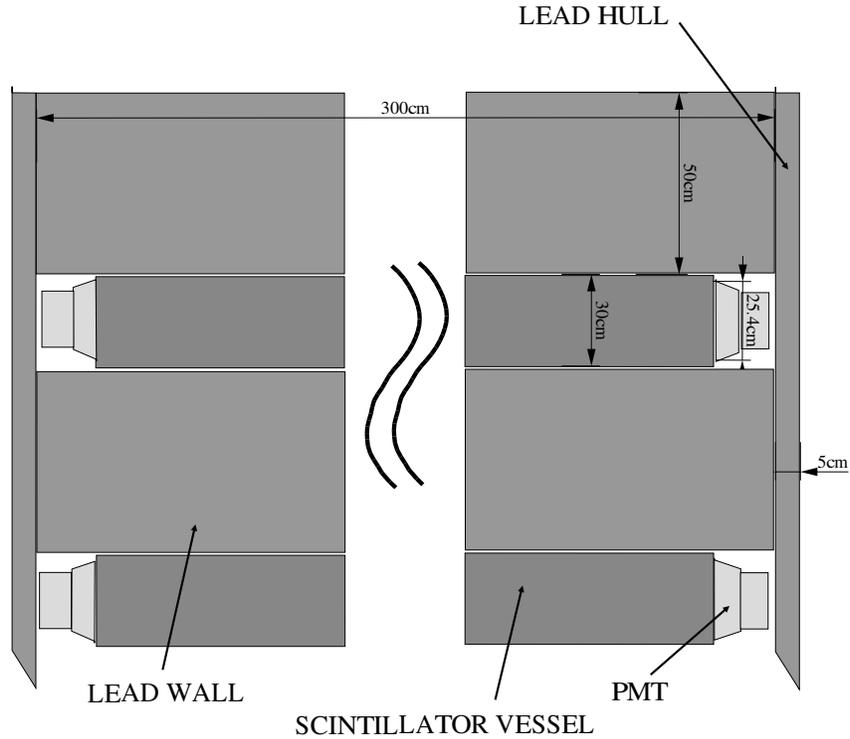}
\caption{Top View of Part of an OMNIS Lead Module\label{f2}}
}
\end{figure*}

Generally, the more fine-grained neutron sources and detection elements are, the higher will be
the neutron detection efficiency, since neutron energy losses and captures before they 
reach the detector will be minimized \cite{psmith}. One important cost overhead though will be 
the data acquisition system; this imposes a practical upper limit of several hundred 
independent scintillator units per $kT$. It is important for the lead to be free of antimony, which
acts as a significant neutron sink in concentrations above $\sim1\%$. Structural stability 
dictates lead walls with thicknesses of at least $\sim10~cm$, interlaced by layers of
scintillator with a thickness determined by the neutron attenuation length in organic
liquids, which is several $cm$. Rectangular acryllic tubes are assumed as scintillator containment 
vessels, which guide the scintillation light via total internal reflection off the acryllic/air 
interface. Their length will be limited by the attenuation length for photons in the scintillator, 
which is about $450~cm$ \cite{paloverde}. The rectangular shape maximally fills out 
the scintillator walls, and can be manufactured for a considerably lower price than comparable 
units with different geometries and materials. For the regions on the tank ends not occupied by 
the photomultiplier tubes, a highly reflective coating of aluminum or $Ti_2O$ is planned. A 
similar design was used for the Palo Verde \cite{paloverde} reactor experiment, which utilized 
acryllic vessels with $0.1\%$ Gd loaded pseudocumene (PC) as a liquid scintillator in a solution 
with $60\%$ mineral oil. The present analysis has assumed a similar material which has a high 
proton concentration ($\frac{H}{C}\approx1.64$), making it a good moderator. However, to 
increase the range over which the thickness of the detector walls can vary, stacking of 
the scintillator vessels in single-, double- or triple-width columns has been considered. To provide 
shielding from external radiation sources and reflect escaping neutrons back into the detector, 
each module is encased in an external lead or iron hull. Figures \ref{f1} and \ref{f2} show a 
schematic design for a $\frac{1}{2}~kT$ lead module. 

\section{DAMOCLES - Model for OMNIS}

\subsection{Basics of Monte Carlo Code Development}

The Monte Carlo code for OMNIS was designed to provide:
\begin{enumerate}
\item Modularity: The detector consists of three principal regions, the scintillator vessels,
the lead walls and the outer hull which have to be simulated in a large number of different
configurations, necessitating easy control of the detector configuration. In particular, 
the code handles the operation of a large number of identical cells.
\item Parameter Control: The development of OMNIS requires the control of all parameters 
regarding geometry, data acquisition and the materials used, including the composition 
of the scintillator and possible sources of radioactive background in all detector components.
That includes the cost of a given detector configuration, which is necessary to optimize cost 
efficiency.
\item Data Flow Control: In order to design the best trigger mechanism, 
extraction of any kind of information about particle positions and energy deposited in the detector
was accommodated.
\item Portability: Since the simulations were performed on both single- and parallel-
processor architectures, the code had to be portable among different platforms.
\end{enumerate}

\subsection{Chronology of Events in the Monte Carlo Simulation}

The treatment of events within the Monte Carlo code follows their chronology. Neutrons are 
first generated with appropriate energies from the spallation following neutrino interactions. 
Each neutron is then tracked through the detector, taking into account scattering reactions 
and eventual capture in one of the components or its escape from the detector. In the former 
case, capture $\gamma$-rays are traced through the detector until their eventual capture or 
escape. Neutrons leaving the detector are not tracked any further. The effects of including 
an additional external reflecting hull were tested beyond that required for shielding out external 
background radiation from the surrounding rock, but the resulting changes were negligible. 
Energy deposited into the scintillator via neutron or $\gamma$-interactions is converted 
into a number of scintillation photons, each of which is then traced until its absorption 
in the liquid, failure to reflect from a surface, or transmission to a photomultiplier tube. 

\subsection{Energy Spectrum of Double-Neutron Events}

In the case of a double-neutron emission event, the available energy is distributed to both 
neutrons \cite{nEspec}, and phase space arguments tend to equalize the energy of both 
neutrons. A Gaussian energy distribution was therefore assumed with the energy for 
each neutron limited to the total available energy, centered at and with a width of half 
the total energy. The influence of the width of the Gaussian on the detection efficiency 
for double-neutron events in lead was investigated. The narrower the Gaussian, the higher 
the efficiency, since the probability for one neutron having very low energy is then small. 
However, the double-neutron efficiency was found not to depend very sensitively on the width; 
the efficiency changed by only $1\%$, if the width is changed to either $10\%$ or $200\%$
of the original value. 

\subsection{Neutron Interactions}

\subsubsection{Neutron Scattering}

The total elastic scattering and absorption cross sections have been approximated using 
polynomials which were fitted to data from the ENDF/B file \cite{endfweb}. Resonances were 
averaged over and not treated in detail, since they are limited to narrow energy ranges. 
At the neutron energies with which OMNIS will be dealing, $\sim1~MeV$, inelastic scattering 
can be neglected. The expansion of the elastic differential scattering cross section in Legendre 
polynomials can be written as
\begin{equation}
\label{dsigmadtheta}
f(cos(\theta ) ,E) = \Sigma _{\ell}(\frac{2\ell+1}{2} a_{\ell}(E) P_{\ell}(cos(\theta ) )),
\end{equation}
where $a_{\ell}$ are the (energy dependent) Legendre coefficients. Step functions 
were used based on the known Legendre coefficients as functions of the energy, and the 
series was terminated at $\ell=5$ for Fe and $\ell=10$ for Pb, beyond which contributions 
are negligible \cite{endfweb}. Both C and H (scintillator) exhibit an s-wave distribution in 
the angular dependence of elastic scattering in the center of mass system for the energy ranges 
considered \cite{endfweb}, thus terminating the series in eq.(\ref{dsigmadtheta}) 
after $\ell=0$. Since its concentration is only $0.1\%$, and its scattering cross section
is comparable to the host material, we neglect scattering from Gd. 

\subsubsection{Neutron Capture}

Neutron capture on Gd is well described by four photons with a total energy of $7.937~MeV$ 
($^{157}Gd$) or $8.536~MeV$ ($^{155}Gd$) \cite{trzcinski96} with individual energies of 
between $50~keV$ and $3.5~MeV$ \cite{capgamma}. Capture on a proton leads to the emission of a $2.2~MeV$ 
photon. Capture on carbon was negligible, as was confirmed by tests. A random $\gamma$-ray of up to 
$4~MeV$ was assumed to be produced by neutron capture on iron or lead. However, our results did not
change at all if this process was left out completely, due to the high absorption cross 
section in Fe or Pb for $\gamma$-rays. For the same reason, any secondary $\gamma$-rays 
produced in inelastic scattering off Pb or Fe were found to be negligible. 

\subsection{$\gamma$-ray Interactions}

The main energy loss channel for $\gamma$-rays in the energy range of interest here, up 
to a few $MeV$, in both the scintillator and the lead or iron, is Compton scattering, although
all possible effects were included. The cross sections for the photoelectric effect, pair 
production, coherent scattering and Compton scattering were obtained from the XCOM database 
\cite{xcom} and fitted for energy bins of $10~keV$ ($1~keV$ for the first $100~keV$ in the scintillator). 
For the angular dependence in the differential cross section for Compton scattering the 
Klein-Nishina equation for photons scattering off electrons was used \cite{jackson}.
Immediate photoelectric absorption was assumed for $\gamma$-rays below an energy of $5~keV$ 
in Fe and Pb and $0.5~keV$ in the scintillator. Below these respective energies, all
interaction channels except photoelectric absorption are negligible, and the 
attenuation length is $<0.6~mm$ in the scintillator, $<0.12~mm$ in lead and $<0.9~mm$ 
in iron \cite{xcom}.

\subsection{Scintillation Photons}
\label{modelphotons}

The liquid scintillator used in the Palo Verde experiment had a light yield of $56\%$ of
that of anthracene \cite{paloverde}, the latter being 16 photons per $1~keV$ \cite{sangster56}. 
The neutron energy conversion efficiency due to the higher charge density caused by recoil 
protons is about $40\%$ of that of the $\gamma$-ray conversion efficiency \cite{price}. 
The phototubes were assumed to detect a given photon with an efficiency of 
$20\%$. Since the tracking of the scintillation photons constitutes the bulk of computing 
time, only $20\%$ of the actual number of photons were simulated, but a 
detection efficiency of $100\%$ was adopted in the phototube. Scintillation photons are created with 
random polarization along either of two possible axes. The absorption cross section in the 
fluid is determined by an exponential law  with an attenuation length of $450~cm$. If the 
incident angle on a surface is larger than the limit for total reflection, the photon is 
reflected with $100\%$ efficiency. The validity of this assumption was confirmed by introducing 
an additional finite absorption probability when a scintillation photon traverses the acryllic 
wall. Even for a transmission of only $95\%$ for every reflection, the efficiency decreases 
by only $1\%$. For all other incident angles, every reflection off the walls was tested for 
absorption, with the reflection probability being a function of polarization and incident 
angle \cite{jackson}. If the photon hits one of the scintillator vessel ends, it is either 
transmitted through the photomultiplier glass face with $98\%$ efficiency or reflected off 
the highly reflective rim surrounding the phototube with an efficiency of $88\%$ for aluminum. 
The latter value was confirmed by measurements conducted at The Ohio State University. 
 
\subsection{Event Identification}

In order to discriminate against noise in the photomultiplier tubes and the
electronics, a threshold was set requiring that one vessel has to register at least five 
photo-electrons on each end within a timebin. The length of a timebin was $100~ns$, in 
accordance with the expected resolution of the data acquisition system. The threshold of 
five photoelectrons was chosen because it is the lowest pulse that can reliably be 
detected in a photomultiplier tube, but the effect of changing that value was studied as well. 
If two such pulses occur in the same vessel within the time-to-amplitude converter time 
window, i.e. if both ends of one vessel therefore fire twice, an event is registered in that 
vessel. Double-neutron events are identified based on two such events registering in 
separate vessels with the first pulse in each event happening within $100~ns$. 

\subsection{Intrinsic Background Radiation}
\label{bgrgeneral}

Assuming that only virgin lead will be used in OMNIS, all major internal background sources
in the lead modules will be elements in the natural decay series or their daughters. 
In lead from the DOE RUN company, a survey of samples exhibited background ranging from 
$<0.001$ to $\sim10~\frac{dps}{kg}$ \cite{oreillyfax}, with most samples towards the lower end of
the range. The dominant background source is $^{210}Pb$ ($T_{\frac{1}{2}}=22.2~yr$) \cite{brodzinski}. 
One of its daughters, the short-lived $^{210}Bi$, emits $1.17~MeV$ $\gamma$-rays \cite{lappandrews}. 
The absence of significant contributions from other isotopes, in particular the $^{226}Th$ 
series, was recently confirmed by CEMRC in Carlsbad \cite{joelwebb}. Bremsstrahlung from 
the various beta-decays can be neglected, because its intensity peaks at energies below 
$\sim0.25~MeV$, where lead absorbs the photons very efficiently. 
\par
The total radioactivity in iron ranges between $0.00$ and $0.17$ decays per second per $kg$ 
above $1~MeV$ \cite{goodmanfax}. Most of the background in iron is due to $^{60}Co$ 
($1.17~MeV$ and $1.33~MeV$, respectively; $60\%$ of the background) and $^{40}K$ 
($1.46~MeV$; $30\%$ of the background). An admixture of $10\%$ $^{226}Th$ ($2.6~MeV$ $\gamma$-ray
from $^{208}Tl$) was assumed for higher energy $\gamma$-ray emissions. 
\par
Another background source is the decay of $^{40}K$ in the glass faces of the photomultiplier 
tubes. The available purity ranges from $1$ to $100$ decays per second per $kg$ per face \cite{photonis}.
\par
 The detected frequency of false $f_F$ events should scale with the background 
rate b as 
\begin{equation}
f_F = k_1 b f_R + k_2 b^2,
\end{equation}
where $k_1$ and $k_2$ are constants depending on the trigger time window and 
$f_R$ is the frequency of real neutron events. The linear term in $b$ enhances 
the detection efficiency by providing a missing second pulse to go with either the
prompt (neutron-) or delayed ($\gamma$-) pulse. The quadratic term in $b$ describes 
coincident double-pulse triggers from the background, in which both pulses within 
the trigger window are caused by background radiation. It is important for the latter 
term to be small at all times, even when there is no supernova, in order to avoid false 
alarms. During the actual supernova pulse, the background contribution is then dominated 
by the linear term. 
\par
The number of background $\gamma$-rays occurring in a time bin, as implemented 
in the code, is Poisson distributed:
\begin{equation}
p_k = \frac{\lambda ^k}{k!} e^{-\lambda },
\end{equation}
where $k$ denotes the resulting number of background events and $\lambda$ is the 
number of events per time bin. 

\section{Efficiency Optimization}

\subsection{Optimization of Detector Dimensions}

The basic philosophy adopted was to optimize the cost efficiency of the OMNIS modules 
rather than the absolute efficiency. Cost efficiency here is defined as efficiency divided by cost 
times $US\$~10^7$, which ranges between $0$ and $2$. The dependence of the detection efficiency 
and cost efficiency on the parameters scintillator vessel dimensions, lead/iron wall 
thickness, hull thickness, photon threshold and neutron energy was investigated.

\subsubsection{The Lead Modules}

The most cost efficient way to detect single-neutron events was found to be a design with 
single scintillator columns with photomultiplier tubes (PMTs) of $10~in$ ($25.4~cm$) diameter.
Double and triple columns of photomultiplier tubes with smaller PMTs were also considered.
However, scintillator vessels with a diameter that is considerably larger than the
attenuation length of both neutrons and $\gamma $-rays dramatically increase the detection 
efficiency and decrease the number of scintillator vessels needed for a given mass of lead. 
This leads to a $50\%$ better cost efficiency for single columns with $25.4~cm$-PMTs compared 
to double columns of $12.7~cm$-PMTs. For double-neutron events, the greater stopping power for 
larger vessels also leads to a higher cost efficiency, more than compensating for the 
detrimental effect of a less fine-grained detector on the detection of two neutrons in 
two separate detectors.
\par
\begin{figure*}
{\centering
\includegraphics[width=10cm]{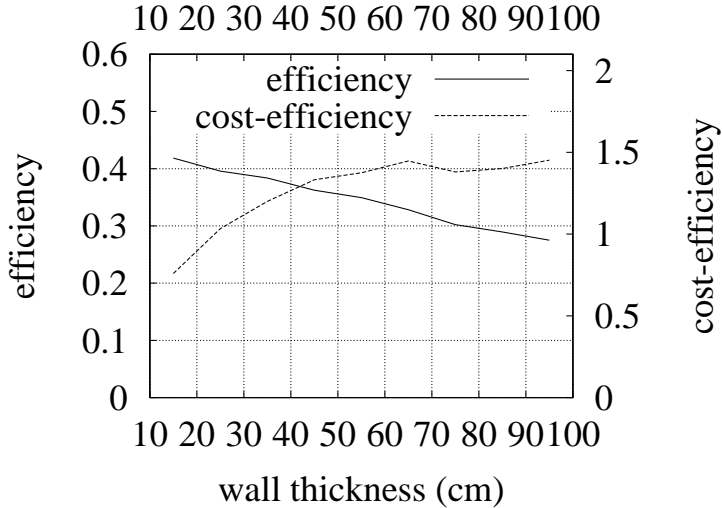}
\caption{Efficiency and cost efficiency vs. lead wall thickness for single-neutron events.\label{lead1}}
}
\end{figure*}

\begin{figure*}
{\centering
\includegraphics[width=10cm]{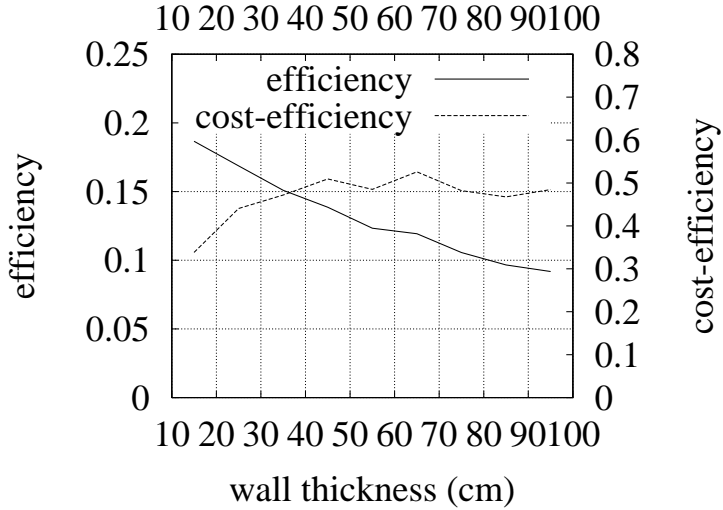}
\caption{Efficiency and cost efficiency vs. lead wall thickness for double-neutron events.\label{lead2}}
}
\end{figure*}

 
The cost efficiency for both single- and double-neutron event detection rises as the lead wall 
thickness increases, then flattens out for a thickness of $50~cm$ (fig.\ref{lead1} and \ref{lead2}), 
although from a pure efficiency-point of view, somewhat thinner lead walls would be preferable.


\begin{figure*}
{\centering
\includegraphics[width=10cm]{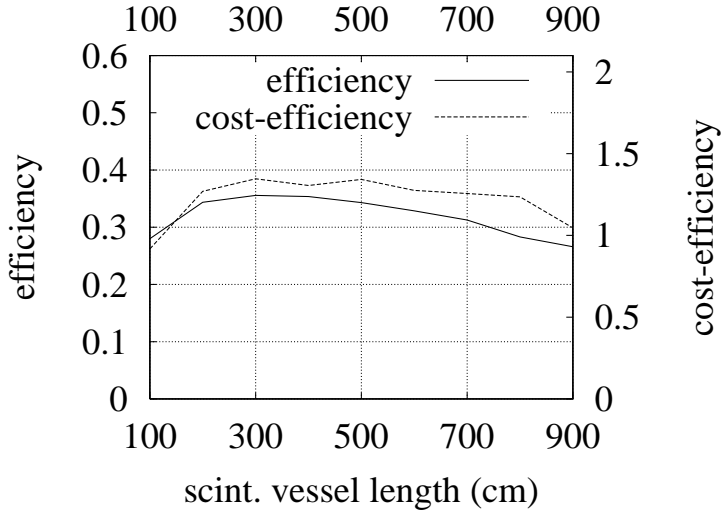}
\caption{Efficiency and cost efficiency vs. scintillator vessel length in the OMNIS lead module for single-neutron events.\label{lead3}}
}
\end{figure*}

\begin{figure*}
{\centering
\includegraphics[width=10cm]{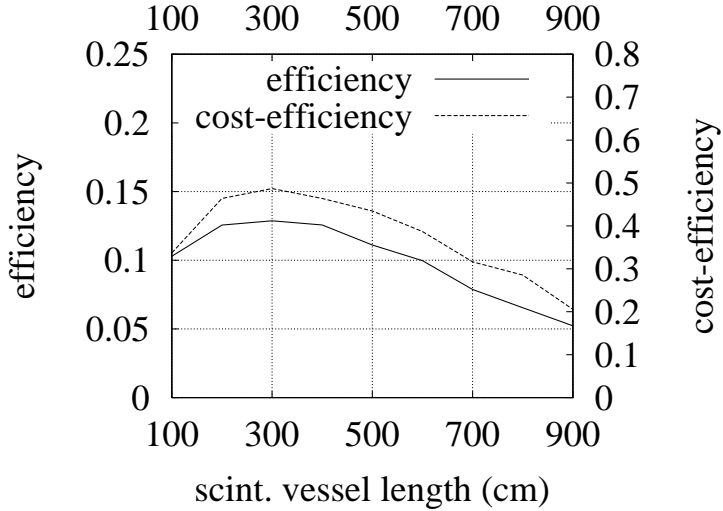}
\caption{Efficiency and cost efficiency vs. scintillator vessel length in the OMNIS lead module for double-neutron events.\label{lead4}}
}
\end{figure*}

As expected from the bulk photon attenuation length in the scintillator, a vessel length of 
$300~cm$ maximizes both efficiency and cost efficiency for single- and especially double-neutron 
event detection (fig. \ref{lead3}, \ref{lead4}). 


The cost efficiency for single-neutron events might be optimized for vessels with widths slightly larger 
than required by the size of the PMTs, the fewer number of PMTs necessary might more than compensate for the
lower efficiency for each single scintillator vessel. There is a slight decrease in the cost efficiency 
with increasing vessel width (by $5-10\%$) and a slight increase (by $\sim3\%$) with vessel 
height, when increased from $30~cm$ to $45~cm$. However, the cost efficiency for two-neutron events 
requires that the vessel dimensions should only exceed the PMT dimensions by a few cm at most, since 
a decrease of the cost efficiency for two-neutron events by $\sim20\%$ is observed when both vessel 
width and height are increased from $30~cm$ to $45~cm$.
\par 
The calculation does not consider the cost for the electronics, although it will be small compared 
to the cost for the scintillator vessels and the PMTs. However, it decreases with an increasing number
of scintillator vessels. We have considered using four PMTs with diameters of $12.7~cm$ instead of 
one $25.4~cm$-PMT per scintillator vessel. This would provide the same fractional coverage of the 
vessel face area, but on top of the approximately four-fold cost for the electronics, one larger PMT 
is less than four times as expensive as four smaller PMTs. 
\par
A thicker lead hull slightly decreases both the detection efficiency and the cost efficiency
of the detector, since more of the neutrons that are produced are lost. We therefore plan to 
make the hull only as thick as necessary to attenuate the external background sources in the WIPP cave. 
\par
\begin{figure*}
{\centering
\includegraphics[width=10cm]{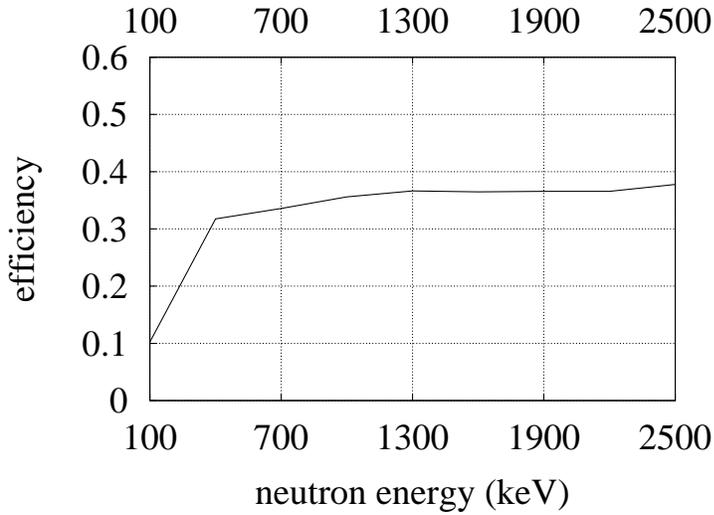}
\caption{Efficiency vs. neutron energy in the OMNIS lead module for single-neutron events.\label{lead9}}
}
\end{figure*}
\begin{figure*}
{\centering
\includegraphics[width=10cm]{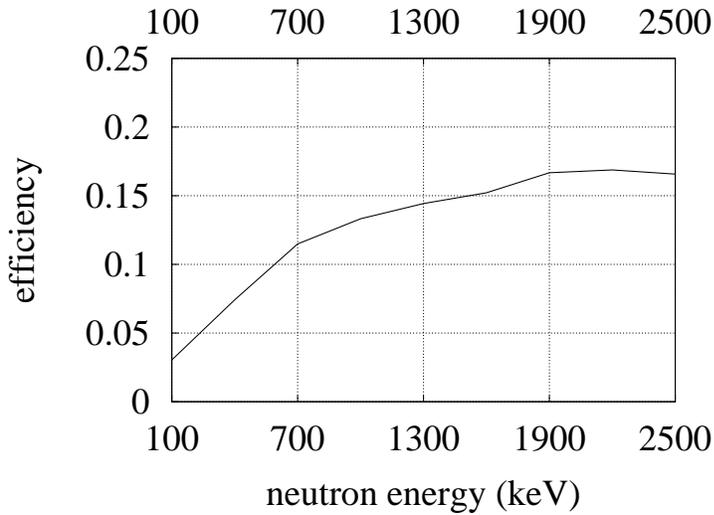}
\caption{Efficiency vs. neutron energy in the OMNIS lead module for double-neutron events.\label{lead10}}
}
\end{figure*}

The detection efficiency for single-neutron events is constant for neutron energies beyond 
a few hundred $keV$ (fig.\ref{lead9}). For double-neutron events, the available energy gets
distributed to two neutrons. Hence, the detection efficiency for those events starts showing
an asymptotic behavior only for energies above $\sim1~MeV$ (fig.\ref{lead10}).


\begin{figure*}
{\centering
\includegraphics[width=10cm]{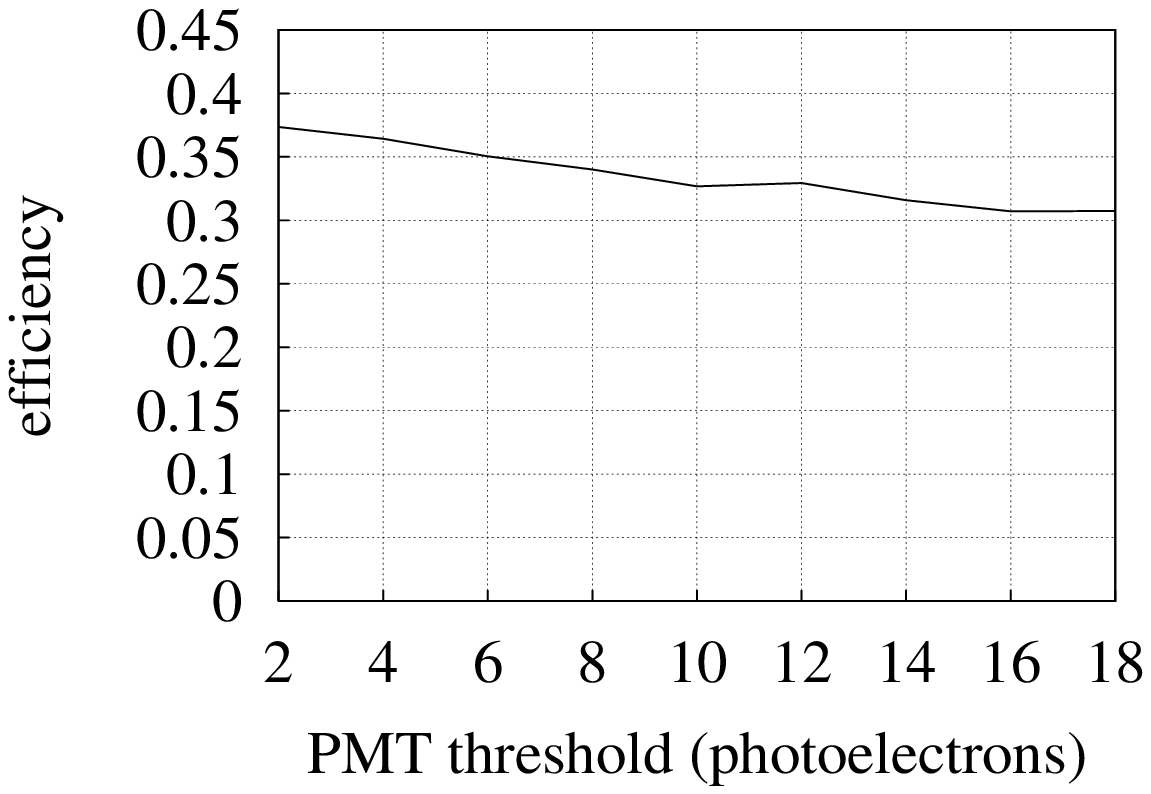}
\caption{Efficiency vs. photoelectron threshold in the OMNIS lead module for single-neutron events.\label{lead11}}
}
\end{figure*}

\begin{figure*}
{\centering
\includegraphics[width=10cm]{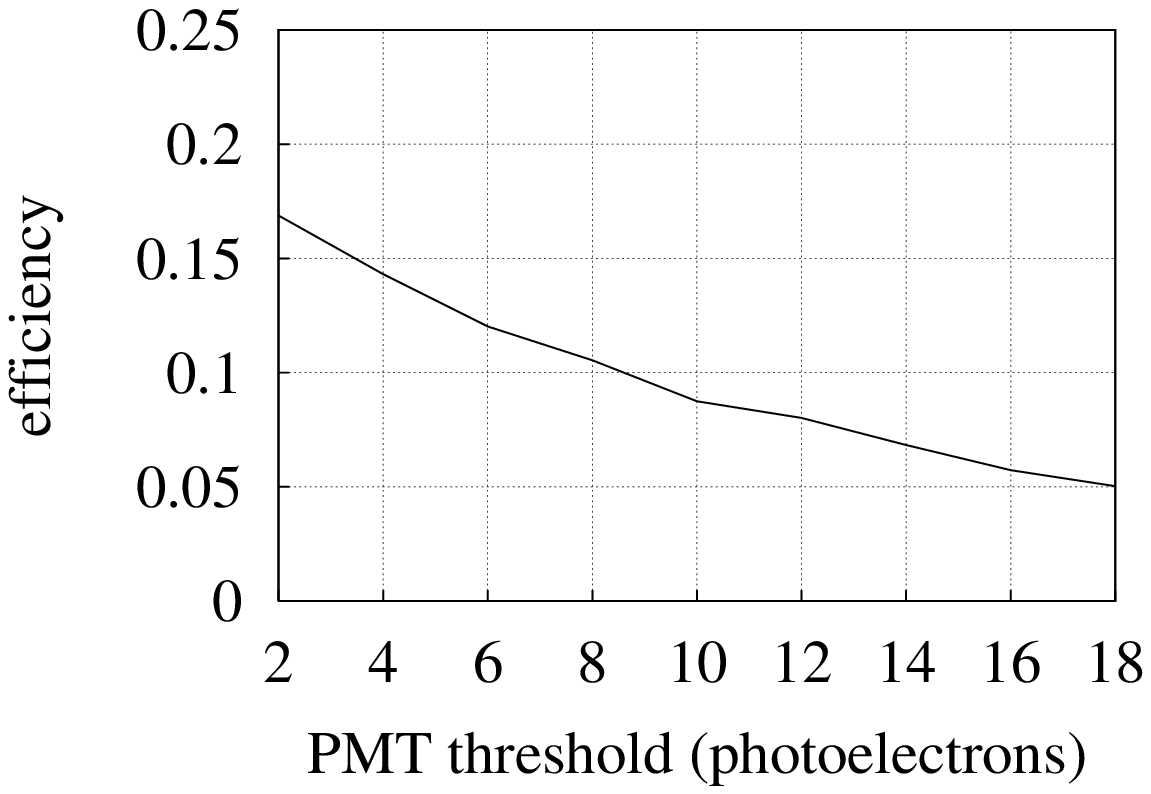}
\caption{Efficiency vs. photoelectron threshold in the OMNIS lead module for double-neutron events.\label{lead12}}
}
\end{figure*}

The detection efficiency is still satisfactory, if the pulse threshold has to be set higher than
the five photoelectrons assumed (fig.\ref{lead11}). However, as expected, the double-neutron 
efficiency decreases more rapidly for higher thresholds (fig.\ref{lead12}). 


\subsubsection{The Iron Modules}

\begin{figure*}
{\centering
\includegraphics[width=10cm]{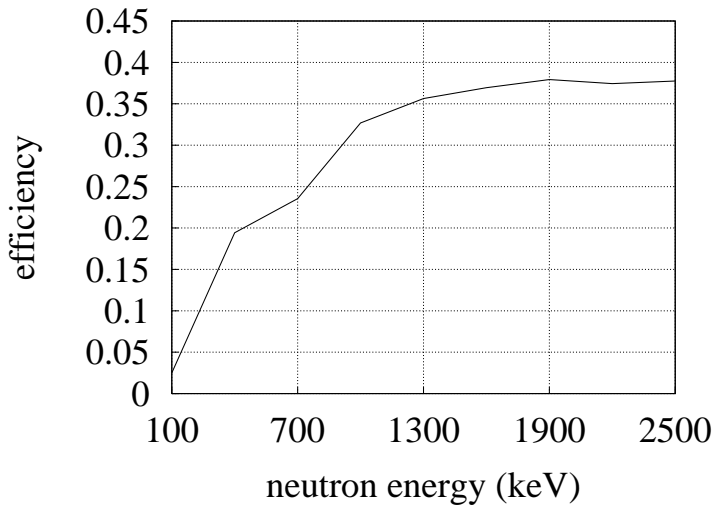}
\caption{Efficiency vs. neutron energy in the OMNIS iron module.\label{steel3}}
}
\end{figure*}

The iron modules have to be optimized only with respect to the cost efficiency for 
single-neutron events. The optimum dimensions of the scintillator containment vessels and
the wall thicknesses are similar to those for the lead modules. As in the case of lead, an iron detector 
with fewer and larger scintillator vessels is both more efficient and cost efficient. The dependence 
of the cost efficiency on the vessel width and height suggests more elongated vessels along the 
vertical direction than in the case of lead. The detection efficiency becomes flat for higher 
neutron energies than in the case of lead (fig.\ref{steel3}), which is due to the fact that 
more energy is lost in an average elastic collision with an iron nucleus. 

\subsubsection{Summary for Optimized Dimensions}

Table \ref{sumdimtable} summarizes the optimized dimensions for $\frac{1}{2}kT$ lead and iron modules,
stacked in single scintillator vessel columns (with photomultiplier tubes with diameters of
$25.4~cm$ on each end).

\begin{table}
\caption{Summary of optimized detector dimensions for the OMNIS lead and iron modules.\label{sumdimtable}}
\begin{center}
\begin{tabular}{|l|r|r|r|r|r|r|}\hline
 Material & wall($cm$) & hull($cm$) & vessel(L($cm$)xW($cm$)xH($cm$)) \\
 \hline\hline
 Pb & 50 & 15 & 300 x 30 x 30 \\
 \hline
 Fe & 50 & 25 & 300 x 30 x 45 \\
 \hline
\end{tabular}
\end{center}
\end{table}

\subsection{Influence of the Background}

\subsubsection{The Lead Modules}

\begin{figure*}
{\centering
\includegraphics[width=10cm]{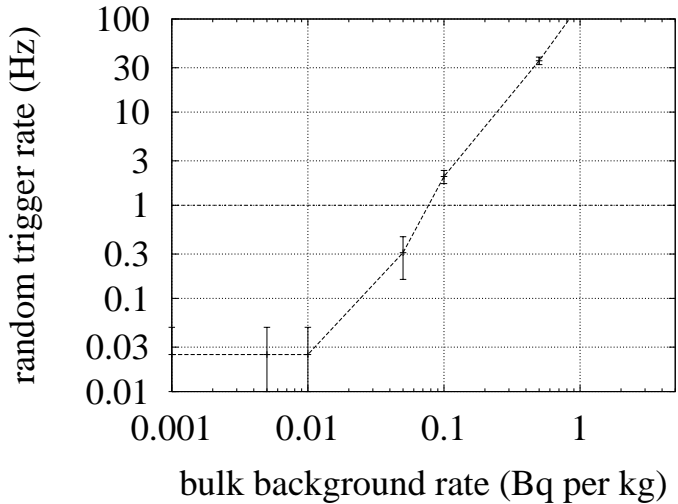}
\caption{False Event Rate per $kT$ due to Bulk Background in OMNIS Pb Modules.\label{pbbulkempty}}
}
\end{figure*}

\begin{figure*}
{\centering
\includegraphics[width=10cm]{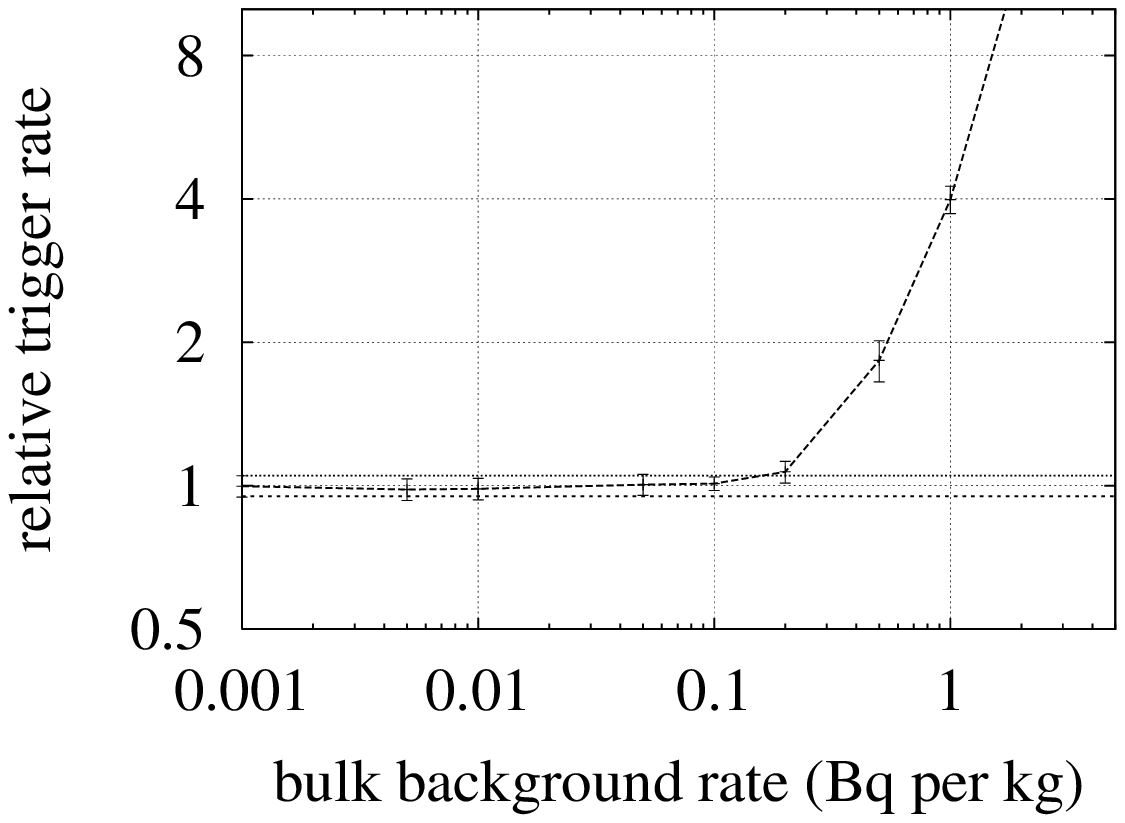}
\caption{Apparent Enhancement of Sparse Neutron Signal due to Bulk Background in OMNIS Pb Modules. The two horizontal lines indicate a $\pm 5\%$ change relative to zero background.\label{pbbulkneutron}}
}
\end{figure*}

The detected false event rate in OMNIS in the absence of emitted neutrons is shown in fig.~\ref{pbbulkempty}.
This confirms radiological analyses~\cite{brodzinski} which show that if the bulk impurity rate in lead 
can be maintained below $0.1~\frac{Bq}{kg}$, then the false event rate in OMNIS will be less than $\sim1$ 
per second. The ability to detect neutrino events at late times after the core bounce is determined by the 
influence of the radioactive background on a sparse neutron signal, with a typical time interval
between two neutrons being long compared to the dead time of the detector, taken to be $80~\mu s$.
The apparent efficiency of a sparse neutron signal relative to the efficiency without background
is shown in fig. \ref{pbbulkneutron}, which shows that a tolerance of $0.1~\frac{Bq}{kg}$ is also
acceptable in this respect. 
\par
\begin{figure*}
{\centering
\includegraphics[width=10cm]{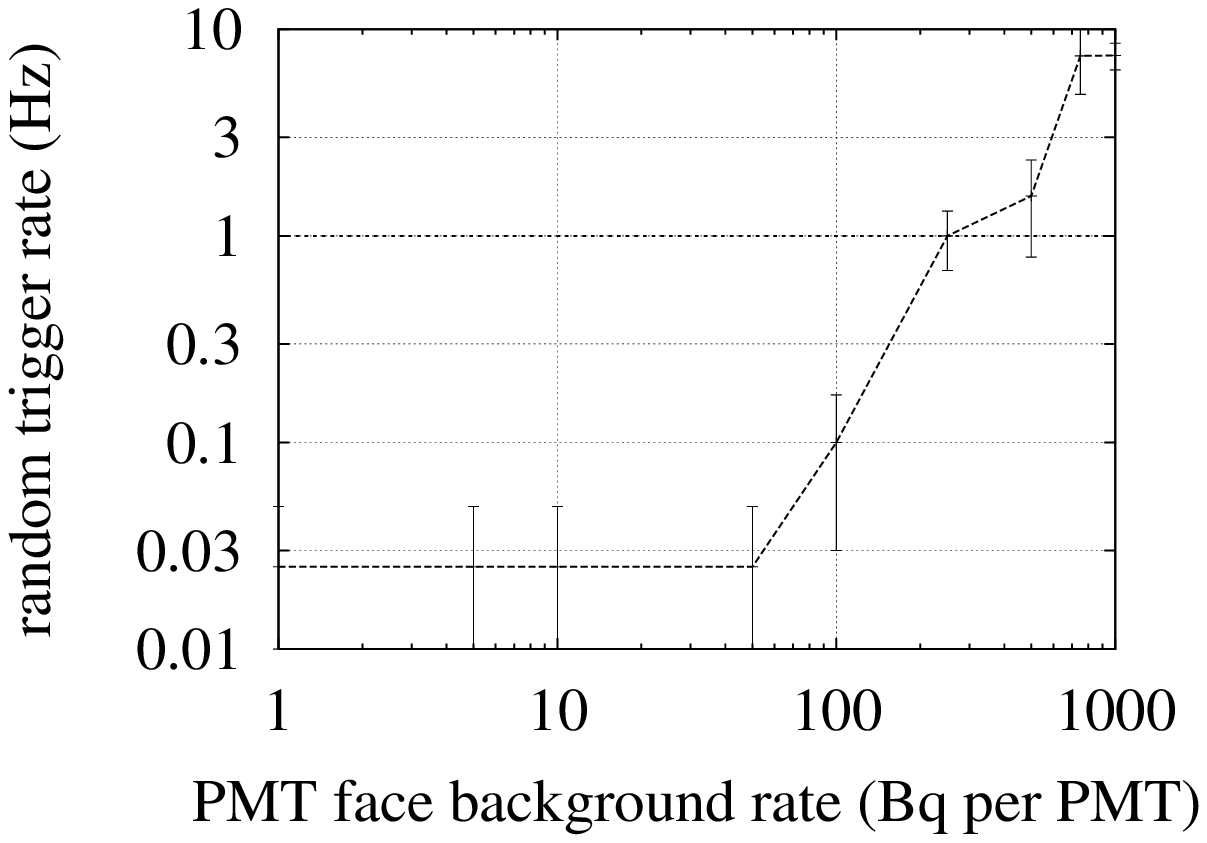}
\caption{False Event Rate per $kT$ due to $^{40}K$ decays in PMT faces in OMNIS Pb Modules.\label{pbfaceempty}}
}
\end{figure*}
\begin{figure*}
{\centering
\includegraphics[width=10cm]{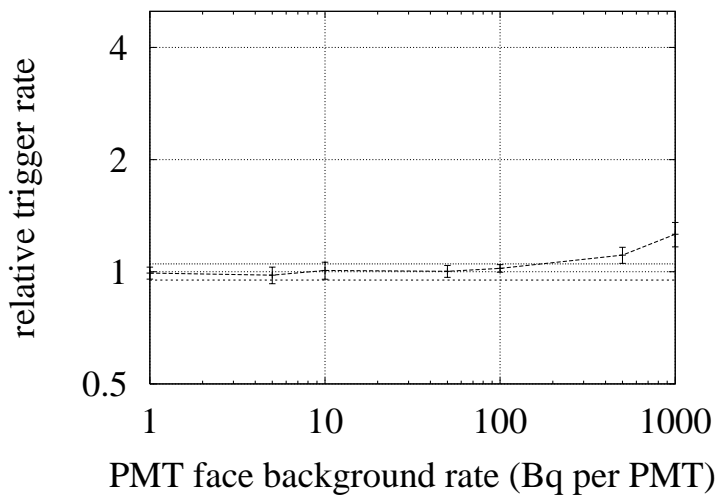}
\caption{Apparent Enhancement of Sparse Neutron Signal due to  $^{40}K$ decays in PMT in OMNIS Pb Modules. The two horizontal lines indicate a $\pm 5\%$ change relative to zero background.\label{pbfaceneutron}}
}
\end{figure*}

Figures \ref{pbfaceempty} and \ref{pbfaceneutron} show the false event rate due to $^{40}K$ decays 
in the photomultiplier glass faces without neutrons and the relative apparent efficiency in the 
presence of a sparse neutron signal. A tolerance limit of $100~Bq$ has been adopted for each PMT. 



\subsubsection{The Iron Modules}

\begin{figure*}
{\centering
\includegraphics[width=10cm]{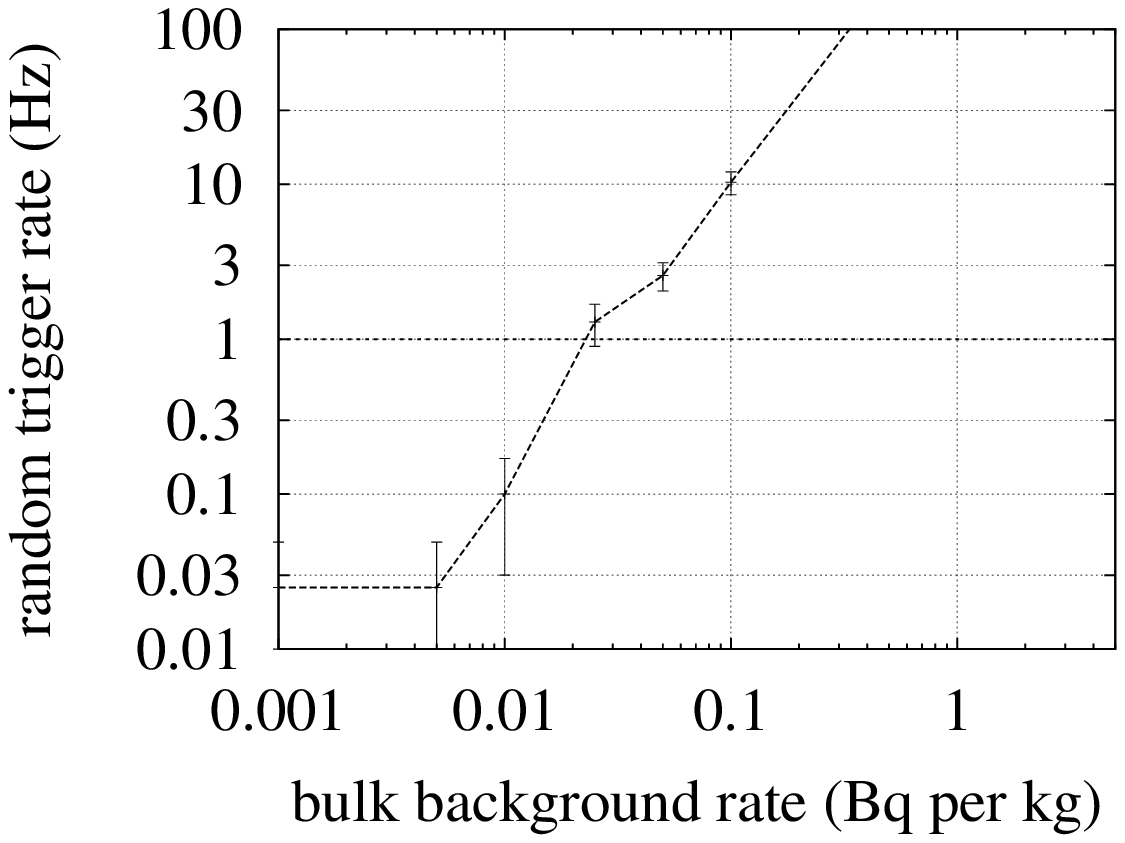}
\caption{False Event Rate per $kT$ due to Bulk Background in OMNIS Fe Modules.\label{febulkempty}}
}
\end{figure*}

\begin{figure*}
{\centering
\includegraphics[width=10cm]{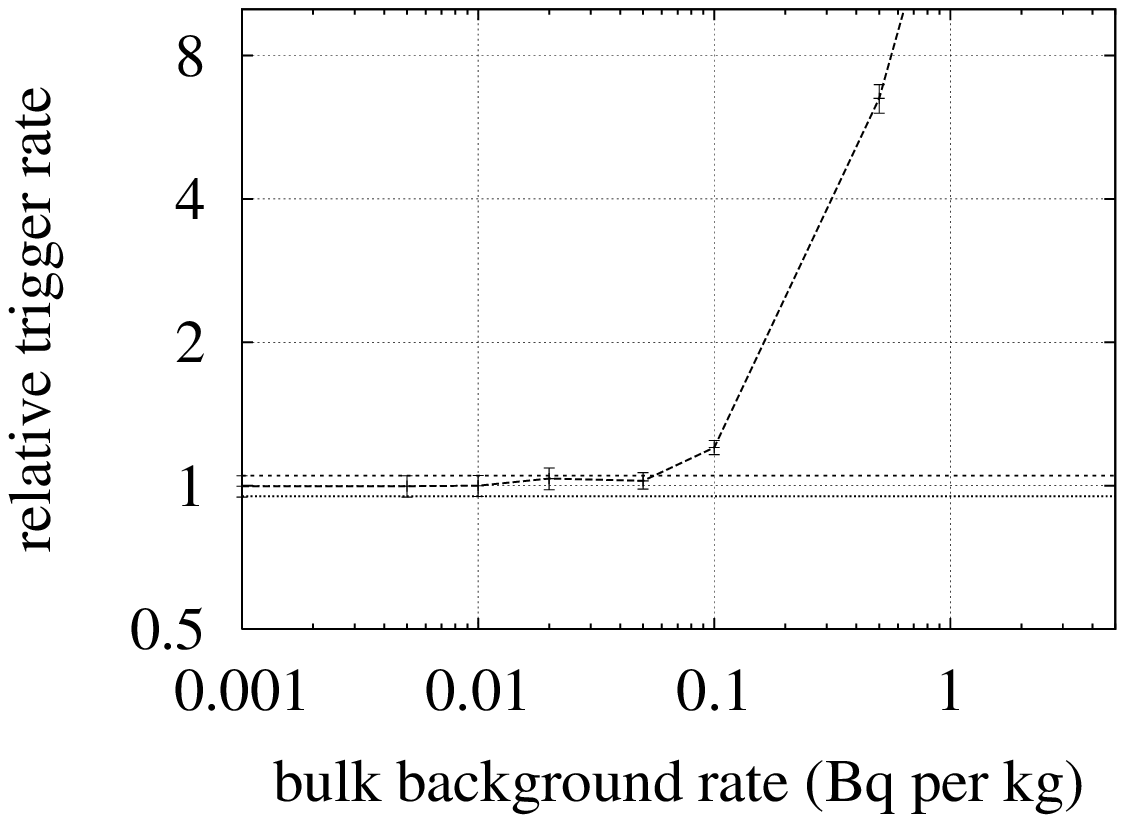}
\caption{Apparent Enhancement of Sparse Neutron Signal due to Bulk Background in OMNIS Fe Modules. The two horizontal lines indicate a $\pm 5\%$ change relative to zero background.\label{febulkneutron}}
}
\end{figure*}

\begin{figure*}
{\centering
\includegraphics[width=10cm]{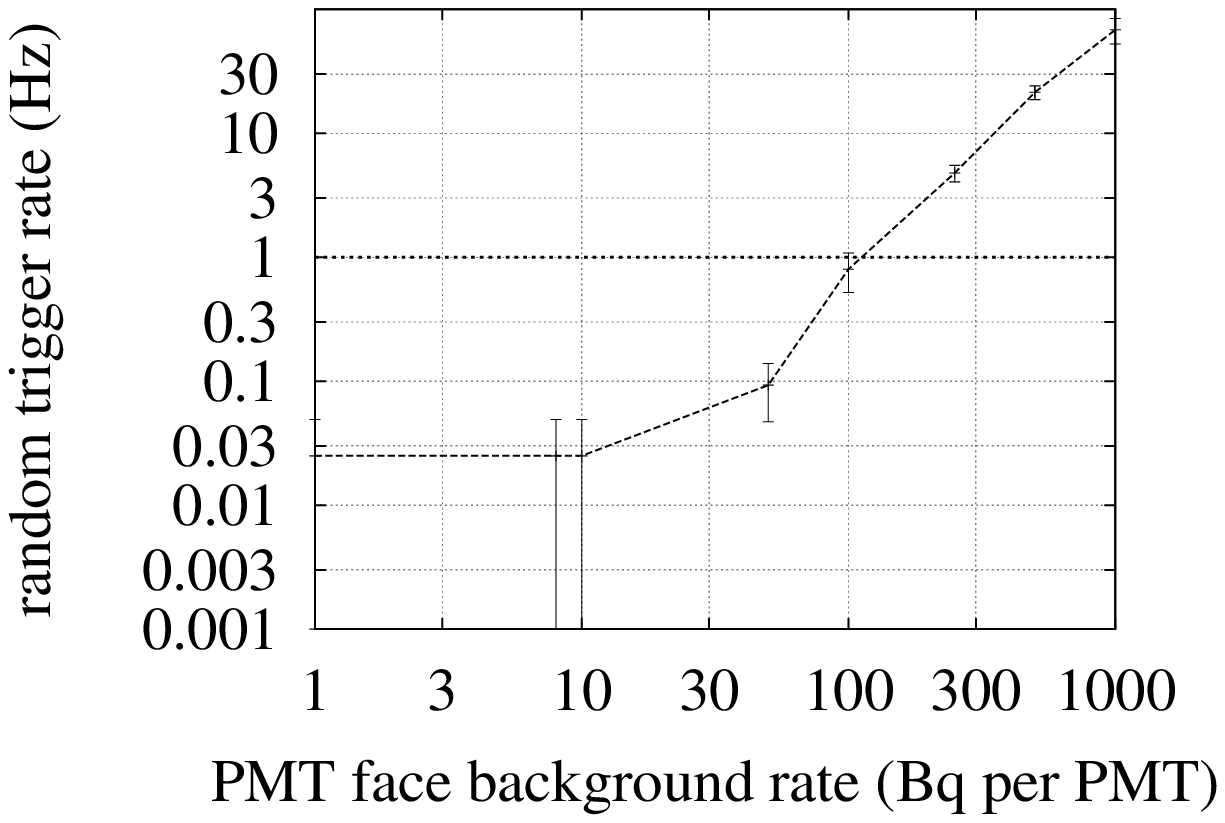}
\caption{False Event Rate per $kT$ due to $^{40}K$ decays in PMT faces in OMNIS Fe Modules.\label{fefaceempty}}
}
\end{figure*}

\begin{figure*}
{\centering
\includegraphics[width=10cm]{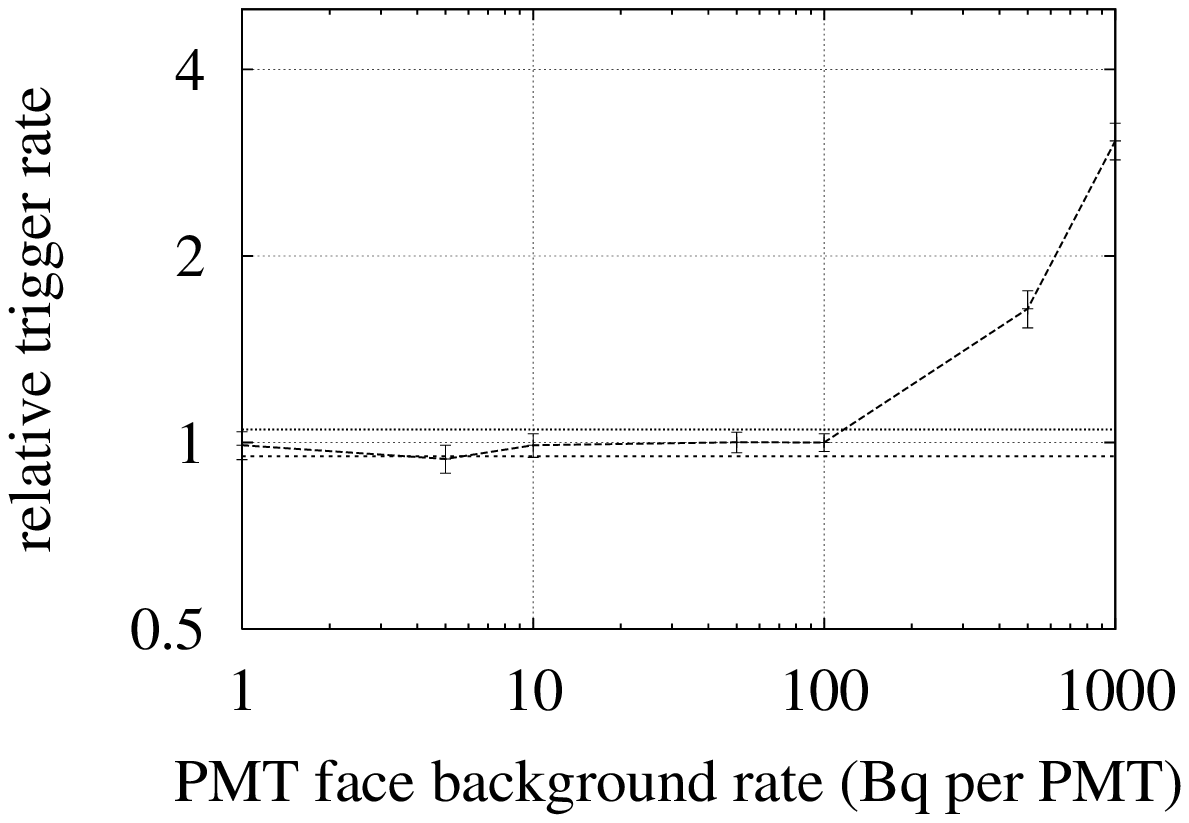}
\caption{Apparent Enhancement of Sparse Neutron Signal due to $^{40}K$ decays in PMT faces in OMNIS Fe Modules. The two horizontal lines indicate a $\pm 5\%$ change relative to zero background.\label{fefaceneutron}}
}
\end{figure*}

Due to the higher energy of the background $\gamma$-rays and their longer attenuation length
in iron, the requirements for its purity have to be more stringent than for lead. The 
resulting tolerance limit for the bulk background in the iron used is $\sim0.05~\frac{Bq}{kg}$ 
(see figures \ref{febulkempty} and \ref{febulkneutron}). This also appears to be consistent 
with radiological analyses \cite{goodmanfax}. The allowed background rate in the photomultiplier 
glass faces is, however, the same as in lead, $100~Bq$ per PMT (see figures \ref{fefaceempty} 
and \ref{fefaceneutron}). 



\subsection{Dependence on Supernova Distance}

If one event strobe is used for an entire detector module, there is a minimum distance of the supernova
for which its dead time becomes critical for reliable event detection. Assuming one detected neutron
event to trigger a strobe with $80~\mu s$ for a lead module of $\frac{1}{2}~kT$ (with 60 scintillator
vessels), dead time losses become significant for supernovae closer than $2~kpc$, which represents only 
a few percent of the Galaxy. Table \ref{distancetable}, generated with a simulated supernova neutrino 
burst \cite{burrows}, shows the number of detected events versus supernova distance. The case without 
dead time is shown for reference. We intend to utilize a sufficiently sophisticated data 
acquisition system such that it can handle the higher neutron liberation rates associated with close 
supernovae in at least one module, thus precluding the excessive loss of high resolution data 
obtainable from a close supernova. 

\begin{table}
 \caption{Number of detected neutron events versus supernova distance for 16 $\frac{1}{2}kT$ Pb modules.
\label{distancetable}}
\begin{center}
\begin{tabular}{|l|r|r|r|r|r|r|r|r|r|}
\hline
Dead Time ($\mu s$) & 0.20 kpc & 0.50 kpc & 1.0 kpc & 2.0 kpc & 4.0 kpc & 8.0 kpc & 16 kpc \\
 \hline\hline
 80 & $0.888\times 10^6$ & $0.293 \times 10^6$ & $95400$ & $26200$ & $6740$ & $1730$ & $440$ \\
 \hline
 0 & $2.86\times 10^6$ & $0.458 \times 10^6$ & $112000$ & $27500$ & $6860$ & $1740$ & $440$ \\
 \hline 
\end{tabular}
\end{center}
\end{table}

\section{Summary}
\label{summary}

Once built, OMNIS will fill in the long perceived need for a large neutral-current supernova neutrino
detector. The neutrino fluxes for the different flavors will be measured using the yields from multiple
threshold processes, single- and double-neutron emission in lead and single-neutron emission 
in iron. The time resolution is only limited by the event rate (and therefore the distance) of 
the supernova. This will reveal unprecedented insight into the core-collapse supernova mechanism 
and might, for the first time ever, open up a window onto the black hole formation mechanism, and 
possibly enable the observation of the different neutrinosphere radii for different neutrino flavors. 
In addition, OMNIS is capable of placing new limits on the allowed neutrino mass - mixing angle 
phase space using time-of-flight methods for different flavors and the relative yields for the different 
neutron liberation channels.
\par 
The generic designs for the OMNIS modules are walls of lead or iron, interlaced with stacks of vessels
containing organic liquid scintillator loaded with Gd. A double-pulse trigger with the capture $\gamma$-ray 
pulse giving the confirmation to the preceding neutron pulse in the same scintillator vessel
was identified as an effective means to discriminate neutron events against the radioactive background.
Using the Monte Carlo code DAMOCLES, developed for OMNIS, optimum configurations
for the detector geometry of the iron and the lead modules were found with respect to the maximum
number of events for a given unit cost. The resulting detection efficiencies for $\frac{1}{2}~kT$ modules 
are $\sim38\%$ for single- and $\sim13\%$ for double-neutrons in lead and $\sim34\%$ for
iron. However, the detection efficiency might improve due to lower possible thresholds when off-line 
analysis of the data is performed. Given the predicted numbers of neutrons liberated, we estimate
a total of $8~kT$ of lead and $4~kT$ of iron  is needed for the OMNIS detector. 
\par
The false trigger rate caused by the radioactive background in the presence or absence of a supernova 
neutrino pulse was investigated with the result that tolerance levels lie within the range found in 
commercially available materials. The criterion was a false rate of $\sim~\frac{1}{sec.}$, low enough 
to still detect a sparse neutron signal. 
\par
The overall neutron trigger rate is below $10~kHz$ for a $\frac{1}{2}~kT$ module and a ``standard'' Galactic supernova 
\cite{burrows,snprofile}, which would enable the use of a common trigger gate for the entire detector. If,
however, a supernova occurs at distances less than $\sim1~kpc$, the typical time between two-neutron events
at the peak of the signal becomes lower than typical dead times of $\sim80~\mu s$, in which case common 
trigger gates for smaller numbers of or even individual photomultiplier tubes become desirable in 
order to measure the peak neutron liberation rate, at least for part of OMNIS.

\section{Acknowledgements}
This work is supported by the National Science Foundation, grant number PHY-9901241 and by The 
Ohio State University. We would especially like to thank the Carlsbad Area Office of the 
Department of Energy and the Waste Isolation Division of the Westinghouse Corporation which
operate the Waste Isolation Pilot Plant for their support. The Monte Carlo simulations were 
supported in part by a grant from the Ohio Supercomputing Center.

\end{document}